\documentclass[prb,12pt]{revtex4}

\usepackage[english]{babel}
\usepackage[intlimits]{amsmath}
\usepackage[dvips]{graphicx}

\begin{document}

\begin{titlepage}
\
\title{Transport of Single Molecules Along the Periodic Parallel Lattices with Coupling}
\author{Evgeny B. Stukalin and Anatoly B. Kolomeisky}

\affiliation{Department of Chemistry, Rice University, Houston, TX 77005-1892 USA}

\begin{abstract}
General discrete one-dimensional stochastic models to describe the transport of single molecules along coupled parallel lattices with period $N$ are developed.  Theoretical analysis that allows to calculate explicitly the steady-state dynamic properties of single molecules, such as mean velocity $V$ and dispersion $D$,  is presented for  $N=1$ and $N=2$ models.  For the systems with $N>2$  exact analytic expressions for the large-time dynamic properties are obtained in the limit of strong coupling between the  lattices  that leads to  dynamic equilibrium between two parallel kinetic pathways.   
\end{abstract}

\maketitle

\end{titlepage}

\section {Introduction}

Successful functioning of biological cells strongly depends on two classes of enzyme molecules, motor proteins and cytoskeleton proteins, that form the basis of the biological transport systems.\cite{bray_book,howard_book}  Motor proteins such as kinesins, dyneins, myosins, RNA and DNA polymerases, etc., operate in cells by transforming the chemical energy of hydrolysis of ATP (or related compounds) into the mechanical work, although the mechanism of this process is still largely unknown.\cite{howard_book} The cytoskeleton proteins, such as microtubules and actin filaments, are rigid multifilament linear polymers that provide tracks for the motion of motor proteins.\cite{bray_book,howard_book,desai97,pollard03} The growth dynamics of these proteins strongly influences their biological functions.\cite{desai97,bray_book,howard_book} One of the most striking properties of growing cytoskeleton proteins is the dynamic instability phenomena that observed both {\it in vitro} and {\it in vivo} in the microtubules\cite{desai97} and in  ParM proteins,\cite{garner04} closely related homologs of the actin filaments.

Recent experimental advances have allowed to determine the dynamics  of motor proteins and the growth of cytoskeleton proteins with a single-molecule precision at different conditions.\cite{howard_book,garner04,kerssemakers03,fujiwara02,asbury03,yildiz03,block05} These experiments suggest that the  complex biochemical transitions and intermediate states  in the motor proteins and cytoskeleton proteins influence their dynamic properties and functions. In order to understand biophysical properties of these proteins  all biochemical pathways and transitions should be taken into account.

One of the most successful approaches to describe the dynamics of motor proteins and growth of cytoskeleton proteins is a method of multi-state discrete chemical kinetic or stochastic models.\cite{qian97,FK,KF00a,KF00b,kolomeisky01,FK2,KF03,stukalin04,stukalin05,kolomeisky05} According to this approach, the motor protein molecule or the tip of the cytoskeleton filament moves between discrete biochemical states, and these transitions are governed by a set of stochastic rates. These biochemical pathways are periodic due to the periodicity in the structure of the cytoskeleton proteins. This method allows to obtain exact analytic expressions for the dynamic properties, such as mean velocity $V$ and dispersion $D$, in terms of the transition rates for systems with arbitrary number of states in the periods.\cite{FK,KF00a,KF00b,kolomeisky01} Another advantage of the discrete stochastic models is the ability to describe the complexity of the underlying biochemical pathways, something that cannot be done by other theoretical methods.\cite{FK,kolomeisky05} Currently, discrete stochastic models are developed, analyzed and applied  for the simple sequential single-chain pathways\cite{FK,FK2,KF03}, for the biochemical pathways with irreversible detachments\cite{KF00a,FK2}, for the lattices with branched states,\cite{KF00a}  for the parallel-chain pathways,\cite{KF00a,kolomeisky01} and for the systems with general waiting-time distributions for transitions between the states.\cite{KF00b}

In order to describe better the complex biological transport processes, in this work we extend the discrete chemical kinetic approach by considering a two-chain model with a direct coupling between parallel lattices at each site, as illustrated in Fig. 1. There are total $2N$ sites in each period of the system, equally divided between the chains. Each site corresponds to a specific biochemical state of the the motor protein or the cytoskeleton protein. The protein molecule can be found on the upper biochemical pathway (chain 0) or it can move  along the lower pathway (chain 1): see Fig. 1. The particle at state $j$ ($j=0,1,\cdots,N-1$) on the chain 0 can make a forward step with the rate $u_{j}$ and a backward step with the rate $w_{j}$ --- see Fig. 1. It also can make a vertical transition  to the state $j$ on the chain 1 with the rate $\gamma_{j}$. Similarly, the particle at state $j$ on the chain 1 can diffuse forward (backward) with the rate $\alpha_{j}$ ($\beta_{j}$), while the transition to the upper channel is given by the rate $\delta_{j}$, as shown in Fig. 1. The distance between identical states in different periods is equal to $d$. This model can be used, for example, to analyze the dynamics of RNA polymerases\cite{block05} or it can describe the dynamic instability of microtubules and ParM proteins.\cite{desai97,garner04}

Our analysis of the periodic coupled parallel-chain discrete stochastic models is based on the Derrida's method\cite{derrida83} that allows to obtain explicit analytic expressions for the stationary-state drift velocity
\begin{equation}
V=V( \{ u_{j},w_{j},\alpha_{j},\beta_{j},\gamma_{j},\delta_{j} \})= \lim_{t \rightarrow \infty} \frac{d}{dt} \langle x(t) \rangle,
\end{equation}
and dispersion (or diffusion constant)
\begin{equation}\label{dispersion}
D=D( \{ u_{j},w_{j},\alpha_{j},\beta_{j},\gamma_{j},\delta_{j} \})= \frac{1}{2}\lim_{t \rightarrow \infty} \frac{d}{dt} \left[ \langle x^{2}(t) \rangle - \langle x(t) \rangle^{2} \right],\end{equation}
where $x(t)$ represents the spatial coordinate of the motor protein or of the tip of the cytoskeleton protein.

The paper is organized as follows. The results for  discrete parallel-chain models in the limit of strong coupling between the pathways are outlined in Section II, while general explicit formulas for dynamic properties for $N=1$ and $N=2$ models  are discussed in Section III. Section IV summarizes and concludes are analysis. The details of mathematical calculations are given in Appendix.

\section{Transport Along Parallel Channels in the Limit of Strong Coupling}

Consider the general periodic coupled two-chain discrete stochastic model as shown in Fig. 1. Let us define $J(j,0)$ and  $J(j,1)$ as probability density currents between the states $j$ and $j+1$ in the channel 0 and 1, correspondingly. Also $J(j,v)$ is a vertical probability density current from the state $j$ in the channel 0 to the state $j$ in the channel 1, assuming that this current is positive in the downward direction (see Fig. 1). The stationary-state conditions and the periodicity of the system require that $J(j,0) + J(j,1)$ to be a constant for any $j=0,1,\cdots,N-1$. Then it leads to the following condition on the vertical currents,
\begin{equation}\label{vertical_currents}
\sum_{j=0}^{N-1} J(j,v)=0.
\end{equation}

Generally, each vertical current is not equal to zero, and  to calculate the dynamic properties of the system analytically is mathematically very hard, except for $N=1$ and $N=2$ models  as will be shown below. However, when the vertical transitions rates $\gamma_{j}$ and $\delta_{j}$ are much larger than other transition rates, one expects that in the limit of large times there will be a ``dynamic'' equilibrium between two channels, i.e., $J(j,v)=0$ at each site. We call this a strong coupling limit. In this case, as shown in Appendix, it is possible to obtain exact analytic expressions for the dynamic properties of the system. Specifically, the equation for velocity is written as
\begin{eqnarray}\label{velocity}
V = \frac{ d \left( 1 - \prod\limits_{j = 0}^{ N-1 } \frac{\tilde{w}_j}{\tilde{u}_j} \right)}{\sum\limits_{j = 0}^{N-1} \left( 1 + \frac{\gamma_j}{\delta_j} \right) r_j},
\end{eqnarray}
and the expression for dispersion is
\begin{eqnarray}\label{dispersion1}
D = \frac{d^2}{N^{2}} \left\{ \frac{1}{ \left[ \sum_{j = 0}^{ N-1 } \left( 1 + \frac{\gamma_j}{\delta_j} \right) r_j \right]^2} \left[ N \sum_{j = 0}^{ N-1 } \tilde{u}_j s_j r_j + A \sum\limits_{j = 0}^{ N-1 } s_j \sum\limits_{i = 0}^{ N-1 } (i+1) \left( 1+ \frac{\gamma_{j+i+1}}{\delta_{j+i+1}} \right) r_{j+i+1} \right]  \right. \nonumber \\
 \left. - A \frac{N+2}{2} \right\},
\end{eqnarray} 
where $ A = NV/d$, and the modified transition rates  are
\begin{eqnarray}
\tilde{u}_j = u_j + \frac{\gamma_j}{\delta_j} \alpha_j, & & \tilde{w}_j = w_j + \frac{\gamma_j}{\delta_j} \beta_j .
\end{eqnarray}
Also, the functions $r_{j}$ and $s_{j}$ are defined as
\begin{equation}\label{rej1}
r_j = \frac{1}{\tilde{u}_j} \left( 1 + \sum\limits_{k=1}^{N-1} \prod\limits_{i=1}^{k} \frac{\tilde{w}_{j+i}}{\tilde{u}_{j+i}} \right), 
\end{equation}
and
\begin{equation}\label{sej1}
s_j = \frac{1}{\tilde{u}_j} \left[ \left( 1 + \frac{\gamma_j}{\delta_j} \right) + \sum\limits_{k=1}^{N-1} \left( 1 + \frac{\gamma_{j-k}}{\delta_{j-k}} \right)  \prod\limits_{i=1}^{k} \frac{\tilde{w}_{j+i-i}}{\tilde{u}_{j-i}} \right]. 
\end{equation}

Comparing these results with the velocity and dispersion for single-channel periodic discrete models,\cite{derrida83,FK} it is interesting to note that the dynamic properties of the two-channel coupled model  can be viewed as a an effective motion along the single pathway with the modified transition rates. This is due to the dynamic equilibrium between two parallel pathways in the strong coupling limit.

\section{Dynamics for $N=1$ and $N=2$ Models with General Coupling}

For the periodic two-chain coupled discrete stochastic models with $N=1$ the dynamics can be calculated explicitly for all parameters. Because of Eq. (\ref{vertical_currents})  the vertical current is always equal to zero at all sites, and the dynamics is the same as in the  strong coupling limit. Then the expressions for the mean velocity and dispersion can be easily obtained  from Eqs. (\ref{velocity}) and (\ref{dispersion1}):
\begin{equation}\label{veln1}
V = d \frac{\gamma (u-w) + \delta (\alpha - \beta)}{\gamma + \delta},
\end{equation}
\begin{equation}\label{dispn1}
D = \frac{d^2}{2}\frac{\gamma (u+w) + \delta (\alpha + \beta)}{\gamma + \delta}.
\end{equation}
It can be seen that both mean velocity and dispersion consist of two terms, corresponding to the motion along the pathways 0 and 1. Here $\frac{\gamma}{\gamma +\delta}$ and $\frac{\delta}{\gamma +\delta}$ are the probabilities to find the particle on the channel 0 and 1, respectively.

Now let us consider  $N=2$ periodic two-chain discrete stochastic models. In this case the dynamic properties can be calculated by solving explicitly a set of Master equations (see Appendix). Then the exact expression for the mean velocity can be written as follows,
\begin{eqnarray}\label{veln2}
V  & =  \frac{d}{\Omega} \left\{ (u_0 u_1 - w_0 w_1) [ (\alpha_0 + \beta_0) \delta_1 + (\alpha_1 + \beta_1) \delta_0 + \delta_0 \delta_1 ] \right. \nonumber  \\  
& + (\alpha_0 \alpha_1 - \beta_0 \beta_1) [ (u_0 + w_0) \gamma_1 + (u_1 + w_1) \gamma_0 + \gamma_0 \gamma_1 ]  \nonumber \\ 
& \left. + (u_0 \alpha_1 - w_0 \beta_1) \delta_0 \gamma_1 + (\alpha_0 u_1 - \beta_0 w_1) \gamma_0 \delta_1 \right\},
\end{eqnarray}
where 
\begin{eqnarray}\label{omega}
\Omega  & = (u_0 + w_0 + \gamma_0) [(\alpha_0 + \beta_0)\delta_1 + (\alpha_1 + \beta_1)\gamma_1] \nonumber \\
& + (u_1 + w_1 + \gamma_1) [(\alpha_0 + \beta_0)\gamma_0 + (\alpha_1 + \beta_1)\delta_0]  \nonumber \\
& + (\alpha_0 + \beta_0 + \delta_0)[(u_0 + w_0)\gamma_1 + (u_1 + w_1)\delta_1]  \nonumber \\
& + (\alpha_1 + \beta_1 + \delta_1)[(u_0 + w_0)\delta_0 + (u_1 + w_1)\gamma_0].
\end{eqnarray}
The corresponding expression for dispersion is quite bulky and  it is presented  in the Appendix.

It is interesting to compare these formulas  with the results from the strong coupling limit. The mean velocity  is given by a simpler equation,
\begin{eqnarray}\label{veln2sc}
V_{SC}=\frac{(u_0 \delta_0 + \alpha_0 \gamma_0) (u_1 \delta_1 + \alpha_1 \gamma_1) -
(w_0 \delta_0 + \beta_0 \gamma_0) (w_0 \delta_0 + \beta_0 \gamma_0)}{(u_0 +
w_0) \delta_0 + (\alpha_0 + \beta_0) \gamma_0)(\gamma_1 + \delta_1) + (u_1 +
w_1) \delta_1 + (\alpha_1 + \beta_1) \gamma_1)(\gamma_0 + \delta_0)}.
\end{eqnarray}
This formula directly corresponds to Eq. (\ref{velocity}).

In order to illustrate the effect of the coupling between the pathways we represent the vertical transition rates in the following form
\begin{equation}
\gamma_{j}(\varepsilon, p)=\gamma_{j}^{0} \varepsilon p, \quad \delta_{j}(\varepsilon,p)=\delta_{j}^{0} \varepsilon (1-p).
\end{equation}
We take $\gamma_{j}^{0}$ and $\delta_{j}^{0}$ as fixed parameters and vary only the parameters $\varepsilon$ ($\varepsilon>0$) and $p$ ($0 \le p \le 1$). The coefficient  $\varepsilon$ gives a measure of the coupling between the channels. When $\varepsilon \gg 1$ we reach the strong coupling limit, while for small values of $\varepsilon$ the coupling is rather weak. The coefficient $p$ reflects the direction of the dominating vertical current. For $p \simeq 0$ the current is mainly upward from the chain 1 into the chain 0, while for  $p \simeq 1$ the vertical current changes the direction (see Fig. 1). The mean velocities for $N=2$ model with different sets of parameters are presented in Fig. 2. It can be seen that the mean velocities do not depend strongly on the coupling between the pathways. Thus the expression for the mean velocity generally can be well approximated by the strong coupling  result.

The situation is different for the diffusion constant. Dispersions for different parameters  are plotted in Fig. 3. It is interesting to note that for a given set of the transition rates dispersion reaches a maximum for low values of $\varepsilon$ and intermediate values of the parameter $p$. When the coupling between the channels is weak  ($\varepsilon=1$) the particle spends most of the time by diffusing in one of the channels before transferring to another one, and dispersion is large. However, for strong coupling ($\varepsilon \rightarrow \infty$) the transitions between the vertical states are very frequent. The particle moves a short distance along the channel before transferring to the other chain, and, as a result, dispersion is low. The situation for low couplings between the channels is very similar to the picture of dynamic instability in microtubules.\cite{howard_book,desai97} Note also that this behavior cannot be observed in $N=1$ models. This suggests that intermediate biochemical states might play a critical role for understanding  dynamic instability in the cytoskeleton proteins.

\section{Summary and Conclusions}

Parallel-chain periodic discrete stochastic models with coupling between the channels are introduced and studied theoretically. The exact analytic expressions for the mean velocity and dispersion are found for the models with arbitrary period size in the limit of strong coupling. This corresponds to the dynamic equilibrium between the pathways and zero currents in the vertical directions at each state.  For the two-chain discrete models with $N=1$ and $N=2$ periods the dynamic properties are calculated explicitly for general conditions. It is found that for weak coupling between the channels dispersion is maximal, while in the dynamic equilibrium regime it is significantly lower. At the same time, the effect of the coupling on the mean velocities is much smaller. Our analysis can be extended to more general parallel-chain discrete stochastic models with more than two different pathways. In addition, it will interesting to to study the dynamics of these models under the influence of external forces.\cite{FK} It is also suggested that these models can be used to investigate the transport of motor proteins and growth dynamics of cytoskeleton proteins.

\section*{Acknowledgments}

The authors would like to acknowledge the support from the Welch Foundation (grant  C-1559), the Alfred P. Sloan foundation (grant  BR-4418)  and the U.S. National Science Foundation (grant CHE-0237105). The authors also are grateful to M.E. Fisher and H. Qian for valuable discussions and encouragements.

\section*{Appendix}

\setcounter{equation}{0}

\renewcommand{\theequation}{\mbox{A\arabic{equation}}}

Let us introduce the probabilities $P_0(l;t)$ and $P_1(l;t)$ of finding the particle at  time $t$ at the position $l=j+Nk$ ($j=0,...N-1$, $k$ are integers) on the lattice 0 and 1, respectively. These probabilities satisfy the following Master equations,
\begin{eqnarray}\label{me1}
\frac{dP_0(j+Nk,t)}{{dt}} =  u_{j-1} P_0(j-1+Nk,t) + w_{j+1} P_0(j+1+Nk,t) + \delta_j P_1(j+Nk,t) \nonumber \\
  - (u_j + w_j + \gamma_j) P_0(j+Nk,t), &  
\end{eqnarray}
\begin{eqnarray}\label{me2}
\frac{dP_1(j+Nk,t)}{{dt}} = \alpha_{j-1} P_1(j-l+Nk,t) + \beta_{j+1} P_1(j+1+Nk,t) + \gamma_j P_0(j+Nk,t)  \nonumber \\
  - (\alpha_j + \beta_j + \delta_j) P_1(j+Nk,t).  
\end{eqnarray}
The conservation of probability requires that 
\begin{eqnarray}
\sum\limits_{k = - \infty}^{ + \infty } \sum\limits_{j = 0}^{N-1} P_0(j+Nk,t) + 
\sum\limits_{l = - \infty}^{ + \infty } \sum\limits_{j = 0}^{N-1} P_1(j+Nk,t)
  = 1
\end{eqnarray}
at all times.

Following the idea of  Derrida, \cite{derrida83} we define two sets of auxiliary functions for each $j = 0,1,\cdots,N-1$ and $i=0,1$.
\begin{equation}
B_i(j,t)= \sum\limits_{k = - \infty}^{ + \infty } {P_i(j+Nk,t)}.  
\end{equation} 
\begin{equation}
C_i(j,t)= \sum\limits_{k = - \infty}^{ + \infty } {(j + Nk) P_i(j+Nk,t)}. 
\end{equation}
Note that the conservation of probability yields 
\begin{equation}\label{norm}
\sum\limits_{j = 0}^{N-1} {B_0(j,t)} + \sum\limits_{j = 0}^{N-1} {B_1(j,t)} = 1.
\end{equation}
Then from the Master equations (\ref{me1}) and (\ref{me2}) we derive
\begin{eqnarray}\label{Be1}
\frac{dB_0(j,t)}{{dt}} =  u_{j-1} B_0(j-1,t) + w_{j+1} B_0(j+1,t) + \delta_j B_1(j,t) - (u_j + w_j + \gamma_j) B_0(j,t),
\end{eqnarray}
\begin{eqnarray}\label{Be2}
\frac{dB_1(j,t)}{{dt}} =  \alpha_{j-1} B_1(j-1,t) + \beta_{j+1} B_1(j+1,t) + \gamma_j B_0(j,t) - (\alpha_j + \beta_j + \delta_j) B_1(j,t).
\end{eqnarray} 
Similar arguments can be used to describe the functions $C_0(j,t)$  and $C_1(j,t)$: \begin{eqnarray}\label{Ce1}
\frac{dC_0(j,t)}{{dt}} = & u_{j-1} C_0(j-1,t) + w_{j+1} C_0(j+1,t) + \delta_j C_1(j,t) - (u_j + w_j + \gamma_j) C_0(j,t) & \nonumber \\
& + u_{j-1} B_0(j-1,t) - w_{j+1} B_0(j+1,t), &
\end{eqnarray}
\begin{eqnarray}\label{Ce2}
\frac{dC_1(j,t)}{{dt}} = &  \alpha_{j-1} C_1(j-1,t) + \beta_{j+1} C_1(j+1,t) + \gamma_j C_0(j,t) - (\alpha_j + \beta_j + \delta_j) C_1(j,t) & \nonumber \\
& + \alpha_{j-1} B_1(j-1,t) - \beta_{j+1} B_1(j+1,t). &
\end{eqnarray}

Again using  Derrida's method,\cite{derrida83} we assume the following stationary-state behavior 
\begin{equation}\label{ansatz}
B_i(j,t) \to b_i(j), {\text{ }}C_i(j,t) \to a_i(j) t + T_i(j) \hspace{0.5cm} (i = 0,1; \ j=0,1,\cdots,N-1).
\end{equation}
At steady state  we have  $dB_i(j,t)/dt = 0$, and Eqs. (\ref{Be1}) and (\ref{Be2}) yield 
\begin{eqnarray}\label{be1}
0 =  u_{j-1} b_0(j-1) + w_{j+1} b_0(j+1) + \delta_j b_1(j) - (u_j + w_j + \gamma_j) b_0(j),
\end{eqnarray}
\begin{eqnarray}\label{be2}
0 =  \alpha_{j-1} b_1(j-1) + \beta_{j+1} b_1(j+1) + \gamma_j b_0(j) - (\alpha_j + \beta_j + \delta_j) b_1(j).
\end{eqnarray}
To determine the coefficients $a_i(j)$ and $T_i(j)$,  Eq. (\ref{ansatz})  is substituted into the asymptotic expressions (\ref{Ce1}) and  (\ref{Ce2}),  producing
\begin{eqnarray}\label{ae1}
0 = u_{j-1} a_0(j-1) + w_{j+1} a_0(j+1) + \delta_j a_1(j) - (u_j + w_j + \gamma_j) a_0(j) ,
\end{eqnarray}
\begin{eqnarray}\label{ae2}
0 = \alpha_{j-1} a_1(j-1) + \beta_{j+1} a_1(j+1) + \gamma_j a_0(j) - (\alpha_j + \beta_j + \delta_j) a_1(j).
\end{eqnarray}
The coefficients $T_i(j)$ satisfy the following equations \begin{eqnarray}\label{te1}
a_0(j) = & u_{j-1} T_0(j-1) + w_{j+1} T_0(j+1) + \delta_j T_1(j) - (u_j + w_j + \gamma_j) T_0(j) & \nonumber \\
& + u_{j-1} b_0(j-1) - w_{j+1} b_0(j+1), &
\end{eqnarray}
\begin{eqnarray}\label{te2}
a_1(j) = &  \alpha_{j-1} T_1(j-1) + \beta_{j+1} T_1(j+1) + \gamma_j T_0(j) - (\alpha_j + \beta_j + \delta_j) T_1(j) & \nonumber \\
& + \alpha_{j-1} T_1(j-1) - \beta_{j+1} T_1(j+1). &
\end{eqnarray}

The strong coupling limit describes the situation when the rates $\delta_{j}$ and $\gamma_{j}$ are much larger than other transition rates. In this case we have 
\begin{eqnarray}\label{equil}
B_1(j,t) \simeq \frac{\gamma_j}{\delta_j} B_0(j,t), & & C_1(j,t) \simeq \frac{\gamma_j}{\delta_j} C_0(j,t).  
\end{eqnarray}
Then, by introducing the modified transition rates
\begin{eqnarray}\label{tilda}
\tilde{u}_j = u_j + \frac{\gamma_j}{\delta_j} \alpha_j, & & \tilde{w}_j = w_j + \frac{\gamma_j}{\delta_j} \beta_j, 
\end{eqnarray}
from Eqs. (\ref{be1}), (\ref{be2}), and (\ref{equil}) we obtain
\begin{eqnarray}\label{be12}
0 =  \tilde{u}_{j-1} b_0(j-1) + \tilde{w}_{j+1} b_0(j+1) - (\tilde{u}_j + \tilde{w}_j) b_0(j).
\end{eqnarray}
At the same time,  Eqs. (\ref{ae1}) - (\ref{te2}), and (\ref{equil}) lead to
\begin{eqnarray}\label{ae12}
0 =  \tilde{u}_{j-1} a_{j-1} + \tilde{w}_{j+1} a_{j+1} - (\tilde{u}_j + \tilde{w}_j) a_{j}.
\end{eqnarray}
\begin{eqnarray}\label{te12}
(1+ \frac{\gamma_j}{\delta_j}) a(j) = & \tilde{u}_{j-1} T_{j-1} + \tilde{w}_{j+1} T_{j+1} - (\tilde{u}_j + \tilde{w}_j) T_{j} & \nonumber \\
& + \tilde{u}_{j-1} b_1(j-1) - w_{j+1} b_1(j+1).
\end{eqnarray}
Here for simplicity we put $ a_0(j) \equiv a_{j}$ and $ T_0(j) \equiv T_{j}$. Following Derrida's solution,\cite{derrida83} we show that $b_0(j) = \Theta_0 r_j$ with
\begin{equation}\label{rej2}
r_j = \frac{1}{\tilde{u}_j} \left( 1 + \sum\limits_{k=1}^{N-1} \prod\limits_{i=1}^{k} \frac{\tilde{w}_{j+i}}{\tilde{u}_{j+i}} \right)
\end{equation}
The coefficient $\Theta_{0}$ can be found from the normalization condition (\ref{norm}) and from Eq. (\ref{equil}),
\begin{equation}
\Theta_0 = \frac{1}{\sum\limits_{j = 0}^{N-1} \left( 1 + \frac{\gamma_j}{\delta_j} \right) r_j}.
\end{equation}
Then, comparing Eq. (\ref{ae12}) and Eq. (\ref{be12}), we  conclude that $ a_j = A b_1(j) $. The coefficient $A$ can be calculated by summing up Eqs. (\ref{te12}) for $ j = 0,1,\cdots, N-1$,\begin{eqnarray}\label{conA}
A = \sum\limits_{j = 0}^{N-1} \left( 1 + \frac{\gamma_j}{\delta_j} \right) a_j =  \sum\limits_{j = 0}^{N-1} ( \tilde{u}_j - \tilde{w}_j ) b_0(j),
\end{eqnarray}
that yields
\begin{eqnarray}\label{defA}
A = \frac{N \left( 1 - \prod\limits_{j=0}^{N-1} \frac{\tilde{w}_j}{\tilde{u}_j} \right)}{\sum\limits_{j = 0}^{N-1} \left( 1 + \frac{\gamma_j}{\delta_j} \right) r_j}.
\end{eqnarray}

To solve Eqs. (\ref{te12}) we define
\begin{equation}\label{defy}
y_j \equiv \tilde{w}_{j+1} T_{j+1} - \tilde{u}_{j} T_{j}.
\end{equation}
Then  Eqs. (\ref{te12}) can be rewritten as
\begin{eqnarray}\label{ye12}
y_{j} - y_{j-1} = \left( 1 + \frac{\gamma_j}{\delta_j} \right) a_j + \tilde{w}_{j+1} b_0(j+1) - \tilde{u}_{j-1} b_0(j-1). 
\end{eqnarray}
The solution of this equation is
\begin{equation}\label{yjsol}
y_j = \tilde{u}_j b_0(j) + \frac{A}{N} \sum\limits_{i=0}^{N-1}(i+1) \left( 1 + \frac{\gamma_{j+i+1}}{\delta_{j+i+1}} \right) b_0 (j+i+1) + \Theta_1,
\end{equation}
where $\Theta_{1}$ is some unknown constant. From Eq. (\ref{defy}) we obtain
\begin{eqnarray}\label{Tjsol}
T_{j} = - \frac{1}{\tilde{u}_{j} \left( 1 - \prod_{j=0}^{N-1} \frac{\tilde{w}_j}{\tilde{u}_j} \right) } \left[ y_j + \sum_{k=1}^{N-1} y_{j+k} \prod_{i=1}^{k} \frac{\tilde{w}_{j+i}}{\tilde{u}_{j+i}} \right]
\end{eqnarray}

Now we can can calculate the mean velocity using the expression for the mean position of the particle
\begin{eqnarray}
<x(t)> = & \frac{d}{N}\sum\limits_{k = - \infty}^{ + \infty } \sum\limits_{j = 0}^{N-1} (j + Nk)[P_0(j+Nk,t) + P_1(j+Nk,t)] =  \frac{d}{N} \sum\limits_{j = 0}^{N-1} [C_0(j,t) + C_1(j,t)] = \nonumber \\
& \frac{d}{N} \sum\limits_{j = 0}^{N-1} \left( 1 + \frac{\gamma_j}{\delta_j} \right) C_0(j,t).
\end{eqnarray}
The velocity is given by the following equation
\begin{eqnarray}
V =  \lim_{t \rightarrow \infty} \frac{d}{dt} <x(t)> = \frac{d}{N} \sum\limits_{j=0}^{N-1} \left( 1 + \frac{\gamma_j}{\delta_j} \right) a_j = \frac{d}{N} A \sum\limits_{j=0}^{N-1} \left( 1 + \frac{\gamma_j}{\delta_j} \right) b_0(j)= \frac{d}{N}A,
\end{eqnarray}
where the function $A$ is given in Eq. (\ref{defA}).

Similar calculations can be performed to determine  dispersion. Starting from
\begin{eqnarray}
<x^2(t)> = \sum\limits_{k = - \infty}^{ + \infty } \sum\limits_{j = 0}^{N-1} (j + Nk)^2[P_0(j+Nk,t) + P_1(j+Nk,t)],
\end{eqnarray}
and utilizing  Eqs. (\ref{me1}), (\ref{me2}) and (\ref{tilda}), we obtain
\begin{eqnarray}
\lim_{t \rightarrow \infty} \frac{d}{dt} <x^2(t)> = 2 \sum\limits_{j=0}^{N-1} (\tilde{u}_j - \tilde{w}_j) C_0(j,t) + \sum\limits_{j=0}^{N-1} (\tilde{u}_j + \tilde{w}_j) B_0(j,t).
\end{eqnarray}
Then, the diffusion constant can be derived from the definition (\ref{dispersion})
\begin{eqnarray}\label{dint}
D = \frac{d^2}{N^2} \left[ \sum\limits_{j=0}^{N-1} (\tilde{u}_j - \tilde{w}_j) T_j + \frac{1}{2} \sum\limits_{j=0}^{N-1} (\tilde{u}_j + \tilde{w}_j) b_1(j) - A  \sum\limits_{j=0}^{N-1} \left( 1 + \frac{\gamma_j}{\delta_j} \right) T_j \right]
\end{eqnarray}
By substituting the expressions for $T_j$ [using Eqs. (\ref{Tjsol}) and (\ref{yjsol})] into Eq. (\ref{dint}), the final expression for dispersion, given in Eq. (\ref{dispersion1}), is obtained. The unknown constant $ \Theta_1 $ cancels out in the final equation for dispersion.

For $N=2$ models with arbitrary coupling between the channels the dynamic properties can be obtained by solving directly Eqs. (\ref{be1}) and (\ref{be2}). The solutions are
\begin{eqnarray}
b_0(0) = \frac{(u_1 + w_1)(\alpha_0 + \beta_0 + \delta_0)\delta_1 + (\alpha_1 + \beta_1)(u_1 + w_1 + \gamma_1)\delta_0}{\Omega}, \nonumber \\
b_0(1) = \frac{(u_0 + w_0)(\alpha_1 + \beta_1 + \delta_1)\delta_0 + (\alpha_0 + \beta_0)(u_0 + w_0 + \gamma_0)\delta_1}{\Omega}, \nonumber \\
b_1(0) = \frac{(\alpha_1 + \beta_1)(u_0 + w_0 + \gamma_0)\gamma_1 + (u_1 + w_1) (\alpha_1 + \beta_1 + \delta_1)\gamma_0 }{\Omega}, \nonumber \\
b_1(1) = \frac{(\alpha_0 + \beta_0)(u_1 + w_1 + \gamma_1)\gamma_0 + (u_0 + w_0) (\alpha_0 + \beta_0 + \delta_0)\gamma_1}{\Omega}. 
\end{eqnarray}
The expressions for $a_{i}(j)$ follow from the relation $a_{i}(j)=A b_{i}(j)$. Similarly, the general solutions of Eqs. (\ref{te1}) and  (\ref{te2}) yield the explicit expressions for $T_{i}(j)$. These expressions are not shown here because they are quite bulky. Finally, the formula for the velocity is given in Eq. (\ref{veln2}), while for dispersion we have
\begin{eqnarray}
D = \frac{d^2}{4} \left \{ (u_0 - w_0) T_0(0) + (u_1 - w_1) T_0(1) + (\alpha_0 - \beta_0) T_1(0) + (\alpha_1 - \beta_1) T_1(1) \right. \nonumber \\
- A \left( T_0(0) + T_0(1) + T_1(0) + T_1(1) \right) \nonumber \\ 
\left. + \frac{1}{2}(u_0 + w_0)b_0(0) + (u_1 + w_1)b_0(1) + (\alpha_0 + \beta_0)b_1(0) + (\alpha_1 + \beta_1)b_1(1) \right \}. 
\end{eqnarray}

\newpage

\noindent {\bf Figure Captions:} \\

\noindent Fig. 1 \quad  General kinetic scheme for the two-chain periodic stochastic model. Both channels have $N$ discrete states per period. The particle on the upper chain 0 moves forward (backward) with the rate $u_j$ ($w_j$), while on the lower chain 1 the corresponding rates  are $\alpha_j$ and $\beta_j$ ( with $j=0,1,\cdots,N-1$). The vertical transitions between the pathways are given by $\gamma_j$ and $\delta_j$.

\vspace{5mm}

\noindent Fig. 2 \quad  Comparison of mean velocities for $N=2$ coupled parallel-chain discrete stochastic  model as a function of the  parameter $ 0 \le p \le 1$ for different values of the parameter $\varepsilon$ (see text for explanations). Solid line corresponds to $\varepsilon=1$, dashed line is for $\varepsilon=20$, while the dotted line describes the strong coupling limit ($\varepsilon \rightarrow \infty$). The values of the horizontal transition rates used for calculations are  $u_0 = \beta_0 = 5$, $w_0 = \alpha_0 = 1$, $u_1 = \beta_1 = 10$, $w_1 = \alpha_1 = 2$. For transitions between the chains we used the following parameters:  $\gamma_0^{0} =1$, $ \delta_0^{0} =\gamma_1^{0} = 2$, and $\delta_1^{0}=0.6$. For all calculations it was assumed that $d=1$.

\vspace{5mm}

\noindent Fig. 3 \quad  Comparison of dispersions for $N=2$ coupled parallel-chain discrete stochastic model  as a function of the parameter $ 0 \le p \le 1$ for different values of the parameter $\varepsilon$ (see text for explanations).  Solid line corresponds to $\varepsilon=1$, dashed line is for $\varepsilon=20$, while the dotted line describes the strong coupling limit ($\varepsilon \rightarrow \infty$). All parameters used for calculations are the same as in Fig. 2.

\newpage

\begin{figure}[ht]
\begin{center}
\vskip 1.5in
\unitlength 1in
\begin{picture}(4.0,4.0)
  \resizebox{3.375in}{1.755in}{\includegraphics{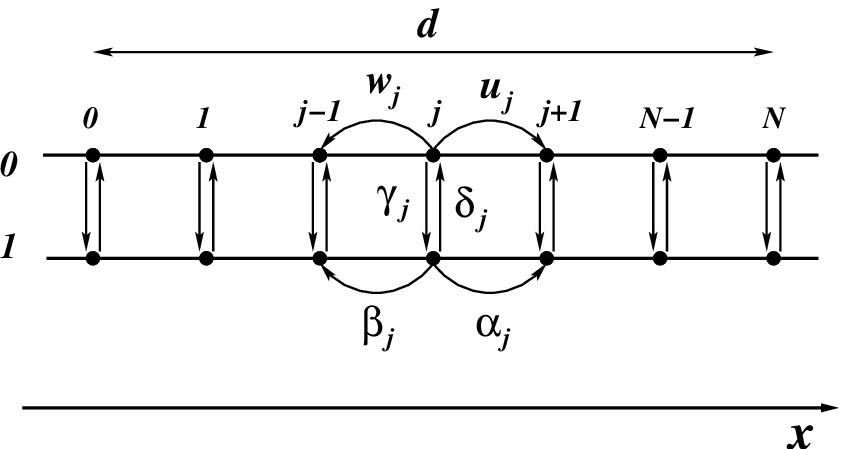}}
\end{picture}
\vskip 3in
 \begin{Large}  \end{Large}
\end{center}
\vskip 3in
\end{figure}

\newpage

\begin{figure}[ht]
\begin{center}
\vskip 1.5in
\unitlength 1in
\begin{picture}(4.0,4.0)
  \resizebox{3.375in}{3.375in}{\includegraphics{v.eps}}
\end{picture}
\vskip 3in
 \begin{Large}  \end{Large}
\end{center}
\vskip 3in
\end{figure}

\newpage

\begin{figure}[ht]
\begin{center}
\vskip 1.5in
\unitlength 1in
\begin{picture}(4.0,4.0)
  \resizebox{3.375in}{3.375in}{\includegraphics{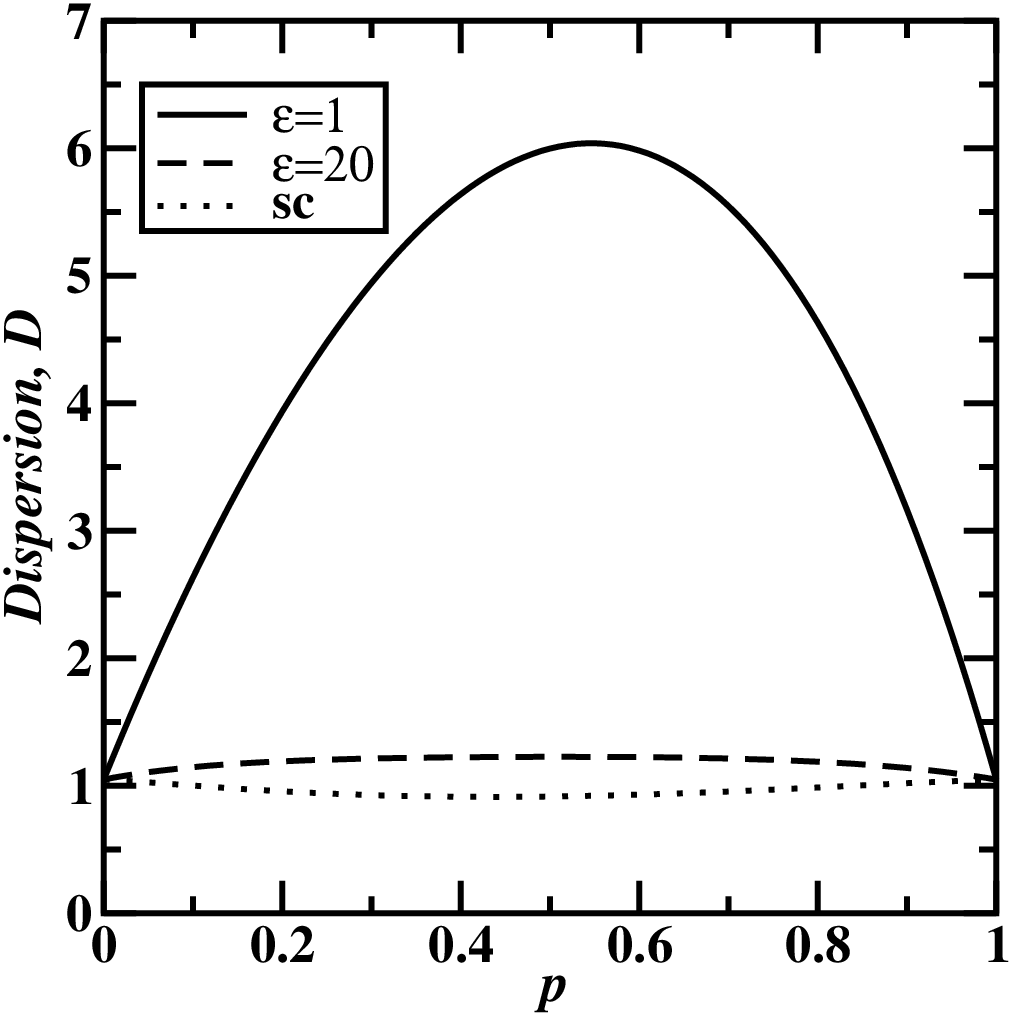}}
\end{picture}
\vskip 3in
 \begin{Large}  \end{Large}
\end{center}
\vskip 3in
\end{figure}


\begin{thebibliography}{99}

\bibitem{bray_book} D. Bray, {\it Cell Movements. From Molecules to Motility}, (Garland Publishing, New York, 2001). 

\bibitem{howard_book} J. Howard, {\it Mechanics of Motor Proteins and Cytoskeleton}, (Sinauer Associates, Sunderland Massachusetts, 2001).

\bibitem{desai97}  A. Desai and T.J. Mitchison, Annu. Rev. Cell. Biol. {\bf 13}, 83 (1997).

\bibitem{pollard03} T. D. Pollard and G.G. Borisy, Cell {\bf 112}, 453 (2003).

\bibitem{garner04} E.C. Garner,  C.S. Campbell and  R.D. Mullins,  Science  {\bf 306} 1021 (2004).

\bibitem{kerssemakers03} J.W.L. Kerssemakers,  M.E. Janson, A. van der Horst, and M. Dogterom,   Appl. Phys. Lett. {\bf 83} 4441 (2003).

\bibitem{fujiwara02} Fujiwara, I., S. Takahashi, H. Tadakuma, T. Funatsu, and S. Ishiwata,  Nature Cell Biol. {\bf 4} 666 (2002).

\bibitem{asbury03} C.L. Asbury, A.N. Fehr  and S.M.  Block, Science {\bf 302} 2130 (2003).
 
\bibitem{yildiz03} A. Yildiz, M. Tomishige, R.D. Vale  and P.R. Selvin,   Science {\bf 302} 676 (2003).

\bibitem{block05} E.A. Abbondanzieri, W.J. Greenleaf, J.W. Shaevitz, R. Landick and S.M. Block, Nature {\bf 438} 460 (2005).

\bibitem{qian97} H. Qian,    Biophys. Chem. {\bf 67} 263 (1997).

\bibitem{FK} M.E. Fisher   and  A.B. Kolomeisky,  Proc. Natl. Acad. Sci. USA {\bf 96} 6597 (1999).

\bibitem{KF00a} A.B. Kolomeisky   and M.E. Fisher, Physica A {\bf 279} 1 (2000).

\bibitem{KF00b} A.B. Kolomeisky  and  M.E. Fisher, J. Chem. Phys. {\bf 113} 10867 (2000).

\bibitem{kolomeisky01} A.B. Kolomeisky, J. Chem. Phys. {\bf 115}, 7253 (2001).

\bibitem{FK2} M.E. Fisher and  A.B.  Kolomeisky,   Proc. Natl. Acad. Sci. USA {\bf 98} 7748 (2001).

\bibitem{KF03} A.B. Kolomeisky   and  M.E. Fisher,   Biophys. J. {\bf 84} 1642 (2003).

\bibitem{stukalin04} E.B. Stukalin  and A.B. Kolomeisky,  J. Chem. Phys. {\bf 121} 1097 (2004).

\bibitem{stukalin05} E.B. Stukalin  and A.B. Kolomeisky,   J. Chem. Phys. {\bf 122} 104903 (2005).

\bibitem{kolomeisky05} A.B. Kolomeisky and H. Phillips III, J. Phys. Condens. Matter {\bf 17} S3887 (2005).

\bibitem{derrida83} B. Derrida, J.Stat. Phys., {\bf 31}, 433 (1983)




\end{thebibliography}
\end{document}